\documentclass{llncs}
\usepackage{xspace}
\usepackage{booktabs}
\usepackage{cite}
\usepackage{color}
\usepackage{url}
\usepackage{paralist}
\usepackage{graphicx}

\newcommand*{\elinda}{\textsc{eLinda}\xspace}

\newcommand*{\linkeddata}{Linked Data\xspace}
\newcommand*{\linkeddataset}{Linked Dataset\xspace}
\newcommand*{\sparql}{\textsc{SPARQL}\xspace}
\newcommand*{\dbpedia}{DBpedia\xspace}
\newcommand*{\yago}{YAGO\xspace}
\newcommand*{\linkedgeodata}{LinkedGeoData\xspace}
\newcommand*{\rdf}{RDF\xspace}
\newcommand{\eat}[1]{}
\def\e#1{\emph{#1}}

\newcommand{\mparagraph}[1]{\vskip 0.5em\noindent\textbf{#1.}\,\,}
\def\partitle#1{\mparagraph{#1}}

\begin{document}
	\title{\huge{\elinda}: Explorer for Linked Data}
	
\author{%
Oren Mishali\inst{1} \and
Tal Yahav\inst{1} \and
Oren Kalinsky\inst{1} \and 
Benny Kimelfeld\inst{1}
}
\institute{Technion -- Israel Institute of Technology}

	\maketitle
	

	\begin{abstract}
		To realize the premise of the Semantic Web towards
		knowledgeable machines, one might often integrate
		an application with emerging RDF
		graphs. Nevertheless, capturing the content of a
		rich and open RDF graph by existing tools requires
		both time and expertise. We demonstrate \elinda---an
		explorer for \linkeddata. The challenge addressed by
		\elinda is that of understanding the rich content of
		a given RDF graph. The core functionality is an
		exploration path, where each step produces a bar
		chart (histogram) that visualizes the distribution
		of classes in a set of nodes (URIs). In turn, each
		bar represents a set of nodes that can be further
		expanded through the bar chart in the path. We allow
		three types of explorations: subclass distribution,
		property distribution, and object distribution for a
		property of choice.
	\end{abstract}
	
	%
	%

	\section{Introduction}
The growing interest in representing information as RDF \linkeddata is
due to the appealing potential of having machines reading and
integrating data from a large and diverse community of resources.
Examples include government efforts such as
Data.gov\footnote{U.S. Government’s open data,
	\url{https://www.data.gov/}} and public initiatives like
\dbpedia.\footnote{\url{http://wiki.dbpedia.org/}} An important
challenge for the research community is to make the data more
accessible for consumption by potential end
users~\cite{journals/semweb/DadzieR11}. In achieving that goal, tools
for effective visualization and exploration of the data play a central
role~\cite{journals/semweb/DadzieR11,journals/corr/BikakisS16}.

The basic principle for effective visual representation of data is
well stated by Shneiderman~\cite{Schneiderman96}: ``Overview first,
zoom and filter, then details-on-demand.'' To that end, a \linkeddataset (or \e{RDF graph}) typically has an associated \emph{ontology}
that describes its semantics in terms of class hierarchies and
properties, and can potentially provide a comprehensive overview of
the underlying data.

The demonstrated tool presented in this paper, \elinda, supports the
exploration and visualization of an RDF graph based on the classes and
properties. Exploration of a dataset is carried over through several
paths, both vertical---down the class hierarchy, and horizontal---via
properties. The visualization is based on bar charts (histograms) that
summarize information about (possibly large) node collections. The
visualization and exploration of \elinda realize a novel formal
framework that we present in Section~\ref{sec:model}.

We drive inspiration from ontology visualization tools, such as
VOWL~\cite{journals/semweb/LohmannNHE15},
FlexViz~\cite{conf/ekaw/FalconerCS10} and
GLOW~\cite{conf/sac/HopRFH12}, that visualize an ontology
independently of the underlying data. Yet, \elinda connects between
ontology elements and their \emph{actual occurrence} in the
dataset. Consequently, during exploration \elinda is able to present the user with ontology elements in a sorted manner, by decreasing significance (i.e., support in
the dataset). This feature is crucial for a user who attempts to make
sense of the data as demonstrated in open and rich datasets.  For
illustration, in \dbpedia the
ontology\footnote{http://mappings.dbpedia.org/server/ontology/classes/}
reports on 49 nine top-level classes, yet almost half of the classes
(22) do not have instances at all.


The ability to dive into the details is a vital requirement for data
visualization tools. A family of tools known as \emph{linked-data
	browsers}~\cite{journals/semweb/DadzieR11} are able to provide
informative insights into the details of a dataset, by supporting the
exploration of resources via properties and relations with other
elements in the \rdf graph. Example tools include
\emph{Marbles}\footnote{http://mes.github.io/marbles/},
\emph{Sig.ma}~\cite{journals/ws/TummarelloCCDDD10}, and \emph{DBPedia
	Mobile}~\cite{conf/www/BeckerB08}. The browsers are focused on
exploring specific \rdf resources and thus present limited, if any,
overview of the data~\cite{journals/semweb/DadzieR11}. In addition,
they often require the user to have prior knowledge of the dataset
(i.e., a starting point), but it is common that users are interested
in finding something useful without previously knowing what they are
looking for~\cite{journals/corr/BikakisS16}. \elinda supports this
mode of exploration, as well as looking into detailed \rdf data with
automatic generation of \sparql queries.

The demonstration will showcase data exploration with \elinda on
several datasets, and using a variety of scenarios and information
needs.

	\def\uris{\mathbf{U}}
\def\lit{\mathbf{L}}
\def\e#1{\emph{#1}}
\def\rdftxt#1{\textsf{#1}}
\def\Qcs{Q_{\mathrm{cs}}}
\def\Qp{Q_{\mathrm{p}}}
\def\Qc{Q_{\mathrm{c}}}
\def\angs#1{\langle #1\rangle}
\def\class{\mathsf{class}}
\def\prop{\mathsf{property}}
\def\set#1{\{#1\}}\
\def\B{\mathbf{B}}
\def\labels{\mathrm{labels}}
\def\init{\mathsf{init}}

\begin{figure*}
	\centering
	\includegraphics[width=1.0\textwidth,height=3in]{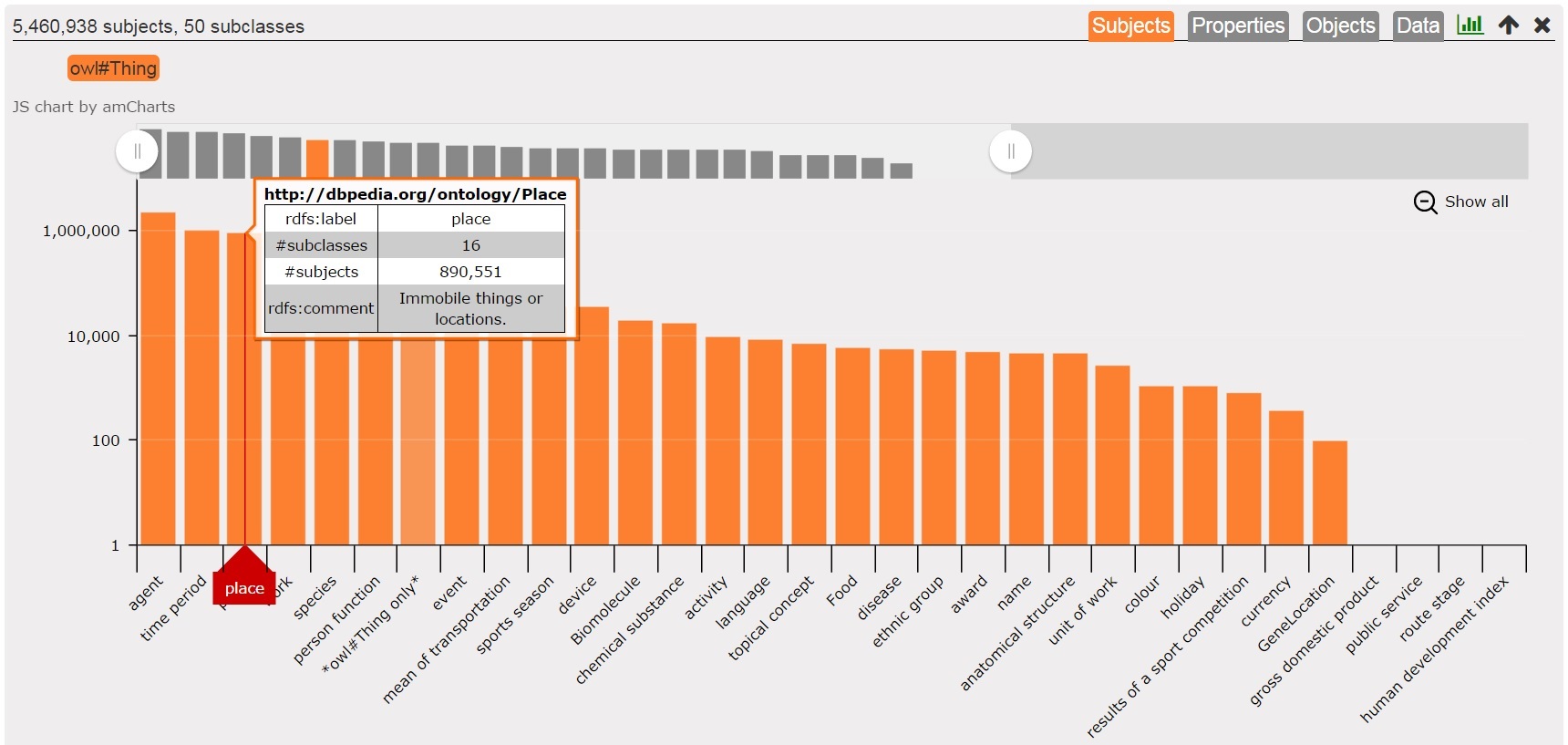}
	\caption{Initial chart in the exploration pane over
		DBPedia}
	\label{fig_initial_pane}
\end{figure*}

\section{Formal Model}\label{sec:model}
In this section we give the formal definition of the data and
interaction model underlying \elinda.

\partitle{RDF graphs} We adopt a standard model of RDF data.
Specifically, we assume (infinite) collections $\uris$ of \e{Unique
	Resource Identifiers} (URIs), and $\lit$ of \e{literals}. An \e{RDF
	triple}, is an element of
$\uris\times\uris\times(\uris\cup\lit)$. An \e{RDF graph} is a finite
collection $G$ of RDF triples. We denote by $\uris(G)$ and $\lit(G)$
the set of URIs and literals, respectively, that occur in $G$. In
the remainder of this section we assume a fixed RDF graph $G$.  A URI
$u$ is said to be \e{of class} $c$ if $(u,\rdftxt{rdf:type},c)\in G$.

\partitle{Bar charts} \elinda enables the visual exploration of $G$ by
means of bar charts, as in Figure~\ref{fig_initial_pane} that are
constructed interactively. Formally, a \e{bar} is a triple
$B=\angs{S,\lambda,t}$ where $S\subseteq\uris(G)$ is a set of URIs,
$\lambda\in\uris(G)$ is the \e{label} of $B$, and $t$ is the \e{type}
of $B$ that can be either $\class$ ($B$ represents URIs that are
associated with some type) or $\prop$ ($B$ represents URIs that are
associated with some property). A \e{bar chart} (or just \e{chart})
$\B$ is associated with a finite set of labels, denoted $\labels(\B)$,
and it maps each label $\lambda\in\labels(\B)$ to a bar with the label
$\lambda$ (i.e., a bar of the form $\angs{S,\lambda,t}$ for some set
$S$); we denote this bar by $\B[\lambda]$.

\partitle{Bar expansion}
A \e{bar expansion} is a function $\eta$ that transforms a given bar
$B$ into a chart $\eta(B)$. \elinda supports three specific bar
expansions $\eta$, each constructing a chart $\eta(B)=\B$ from a given
bar $B=\angs{S,\lambda,t}$.
\begin{itemize}
	\item \textit{Subclass expansion:}\,\, This expansion is enabled when
	$t=\class$. Then, $\labels(\B)$ consists of all $\tau$ such that $G$
	contains $(\tau,\rdftxt{rdfs:subClassOf},\lambda)$, and
	$\B[\tau]=\angs{T,\tau,\class}$ where $T$ consists of all $s\in
	S$ of class $\tau$.
	\item \textit{Property expansion:}\,\, This expansion is enabled when
	$t=\class$. Then, $\labels(\B)$ consists of all $\pi$ where $G$
	contains $(s,\pi,o)$ for some $s\in S$ and $o$, and
	$\B[\pi]=\angs{T,\pi,\prop}$ where $T$ is the set of all the $s\in S$ with
	the property $\pi$ (that is, $(s,\pi,o)\in G$ for some $o$).
	\item \textit{Object expansion:}\,\, This expansion is enabled when
	$t=\prop$. Then, $\labels(\B)$ consists of all $\tau$ such that $G$
	contains $(s,\lambda,o)$ for $s\in S$ and objects $o$ of class
	$\tau$.  The set $\B[\tau]$ is $\angs{T,\tau,\class}$ where $T$
	consists of all the $o$ of class $\tau$ with $(s,\lambda,o)$ for
	some $s\in S$.       
\end{itemize}

In addition, we allow for a \e{filter} operation that is associated
with a condition on URIs, and removes from each bar $B$ the URIs that
violates the condition.

The property and object expansions are defined above for \e{outgoing}
properties, that is, the URIs of $S$ play the roles of the
\e{subjects}. We similarly define the \e{incoming} versions, where the
URIs of $S$ play the roles of the \e{objects}.

\partitle{Exploration}
Finally, \elinda enables the exploration of $G$ by enabling the user
to construct a list of charts in sequence, each exploring a bar of the
previous chart. The exploration begins with a predefined \e{initial
	chart} that we denote by $\B_0$. In our implementation this bar has
the form $\eta(B)$ where $\eta$ is the subclass expansion and
$B=\angs{S,\tau,\class}$ is the bar with $\tau$ being a predefined
type and $S$ consisting of all $s$ with
$(s,\rdftxt{rdf:type},\tau)\in G$. A sensible choice of $\tau$ is a
general type such as $\rdftxt{owl:Thing}$.

More formally, by \e{exploration} we refer to a sequence of the form
$(\lambda_1,\eta_1)\mapsto\B_1\,,\,(\lambda_2,\eta_2)\mapsto\B_2\,,\,\dots\,,\,(\lambda_{m},\eta_{m})\mapsto\B_m$
where each $\B_i$ is obtained by selecting a bar from $\B_{i-1}$,
namely the one with the label $\lambda_i$, and applying the expansion
$\eta_i$ to that bar. (Recall that $\B_0$ is the predefined initial
chart.) In notation we have the following for all $i=1,\dots,m$:
(a) $\lambda_i\in\labels(\B_{i-1})$; 
(b) $\eta_i$ is applicable to $\B_{i-1}[\lambda_i]$;
and (c) $\B_{i}=\eta_i(\B_{i-1}[\lambda_i])$.
As a feature, \elinda enables the user to generate SPARQL code to
extract each of the bars along the exploration.



	\section{User Interface}\label{sec:ui}

\begin{figure}[t]
	\centering
	\includegraphics[width=0.48\textwidth]{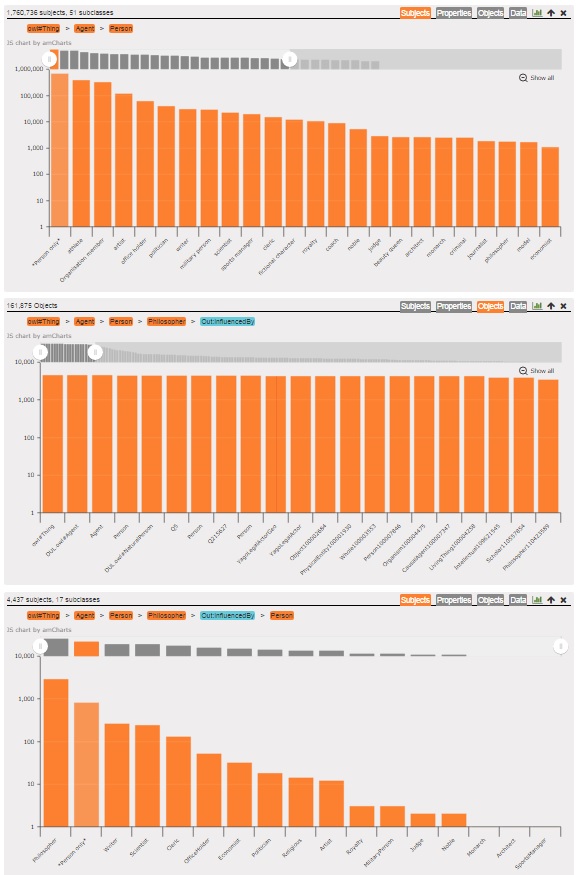}
	\caption{Three exploration panes over \dbpedia{}: the
		\rdftxt{Person} class, the \rdftxt{Philosopher} class, and
		persons influencing philosophers. The colored
		breadcrumb trails indicate the exploration path.}
	\label{fig_panes}
\end{figure}
In this section we explain how the interaction model of the previous section is materialized in \elinda's user interface. We start with basic concepts regarding the operation of the tool.

\subsection{Operational context}
\elinda is implemented as a single-page web application within a web browser. The interface provides a setting form that allows a user to point the tool to an online \sparql endpoint such as \dbpedia, \yago, or \linkedgeodata.\footnote{The current implementation assumes Virtuoso endpoints.}
During its operation, user requests are translated into numerous \sparql queries that are sent to the server in real time, asking for data to visualize.
The very first queries present the user with general statistics about the dataset such as the total number of RDF triples, and the number of classes the dataset has. The exploration however is totally \e{visual}, and does not require knowledge of the \sparql query language.

\partitle{Requirements from the dataset}
A \linkeddataset is usually associated with an \e{ontology} that defines a class type hierarchy and properties. \elinda offers an ontology-based exploration experience, while trying to make minimal assumptions on the structure of the target dataset. For example, even datasets with no class hierarchy at all may be browsed with \elinda however in a limited fashion. Yet the full power of the tool is exploited for datasets that define a class type hierarchy using the standard properties \rdftxt{owl:Class} (or \rdftxt{rdfs:Class}) and \rdftxt{rdfs:subClassOf}. Such datasets will be better ``explorable" since the user will be able to navigate down the class hierarchy while searching for data of interest. With regard to properties, \elinda does not assume their declaration with \rdftxt{rdf:Property} or alike, since they are totally inferred from the actual data triples as explained in Section \ref{sec:properties}. \elinda makes extensive use of standard \rdftxt{rdfs:label} properties, that if exist provide the user with short and meaningful textual labels that are attached to visualized elements.

\subsection{Navigating with the subclass expansion}
Exploration with \elinda is effectively performed by constructing a sequence of \e{tabbed panes}. An example pane for the \dbpedia dataset is presented in Figure \ref{fig_initial_pane}. Each pane visualizes data related to a set of subjects (instances) $S$ from several different perspectives. All subjects in $S$ are of the same type $T$.\footnote{Note that $S$ does not necessarily include \e{all} instances of $T$ as we will explain later on.} When pointing \elinda to a new dataset an initial pane is shown, and during the exploration the user may open additional panes one beneath the other. The initial pane in Figure \ref{fig_initial_pane} refers to all \dbpedia subjects of type \rdftxt{owl:Thing} (the root class).\footnote{We also handle the case of datasets with not root class, as found in \linkedgeodata.} The upper left corner of a pane shows basic statistics: the total number of instances (i.e., $|S|$), and the number of direct and indirect subclasses that class type $T$ has.

\partitle{Subclass chart}
By default, a pane is opened with a bar chart showing the \e{distribution} of instances in $S$ among the subclasses of $T$ (the result of a \e{subclass expansion}).
Each bar represents instances of a specific subclass, and the height of the bar is proportional to the number of instances. The bars are sorted by decreasing height. Hovering a bar reveals a pop-up box with basic information, as shown in the figure for \rdftxt{Agent}, the second largest \dbpedia class, with more than 2 million instances, 5 direct subclasses, and 277 subclasses in total. To facilitate the visualization of a large number of bars, only a subset of the bars is initially shown. A widget located at the top of the chart allows to control visible part of the chart.

\partitle{Class navigation}
A user interested in exploring a particular instance set, or in getting an overview of the dataset, can navigate down the class hierarchy by clicking a bar, what causes a new pane to be opened below, with a new chart showing its direct subclasses. For example, a user wishes to explore instances of type \rdftxt{Philosopher} should open three additional panes through the path \rdftxt{owl:Thing} $\rightarrow$ \rdftxt{Agent} $\rightarrow$ \rdftxt{Person} $\rightarrow$ \rdftxt{Philosopher}. Often, locating the position of a desired class type using top-down navigation may be challenging. For example, it may be difficult for a user to infer that class \rdftxt{Philosopher} is located under \rdftxt{Agent $\rightarrow$ Person}. For such circumstances, \elinda provides an autocomplete search box for locating class types, based on a list that is populated by collecting all subjects in the dataset of type \rdftxt{owl:Class} or \rdftxt{rdfs:Class}. Selecting a class that way, immediately opens the associated pane without the need to drill down.

\subsection{Applying the property expansion}
\label{sec:properties}
Within a pane, switching to the \e{Property Data} tab shows a second chart with a comprehensive overview about properties featured by instances in $S$ (the result of a \e{property expansion}). To calculate this property data, \elinda does not refer to ontology elements such as \rdftxt{rdf:Property} that may not be available in the dataset or incomplete. Instead, it aggregates \e{all} properties found within instances in $S$. The bar chart here shows how the instances are distributed among the properties found. A specific bar corresponds to those instances in $S$ that share a property $p$. The height of a bar is proportional to the
property's \e{coverage}---the percentage of instances that feature
the property in $S$. (Again, bars are sorted by decreasing height.)

The number of possible properties for a given set $S$ may be very large and thus difficult for the user to consume. For example, in \dbpedia there are nearly 40,000 instances of type \rdftxt{Politician}, that feature 1,482 different properties altogether. We enable the user to restrict to significant properties by filtering out properties with a coverage lower than a threshold. In the \rdftxt{Politician} example, only 38 properties that cross the default coverage threshold of 20\% are shown. The user may adjust the threshold and reveal more properties if needed.

\partitle{Browse instance data}
The aforementioned property chart provides an overview, yet in some cases it may be desirable to view actual instance data. This is facilitated by another UI element, a \e{data table}, that presents data in a tabular format. Each bar in the property chart that is selected by the user is added as a new column in the table. The column is then filled-in with actual values that are fetched from the dataset. For example, by selecting the property bars \rdftxt{birthPlace} and \rdftxt{influencedBy} for instances of type \rdftxt{Philosopher}, we can see in the table where each philosopher was born, and by whom he or she was influenced. In addition, the table exposes the \sparql query it was generated from, allowing a user to retrieve the corresponding data.

\partitle{Ingoing properties}
The default property chart relates to \e{outgoing} properties that ``leave" instances of $S$. Similarly, a user may explore properties that ``enter" those instances by switching to an \e{ingoing} property chart. For type \rdftxt{Philosopher}, 9 ingoing properties that cross the 20\% coverage threshold are shown, such as \rdftxt{author} that connects between different works to philosophers who authored them.

\partitle{Data filters}
A \e{data filter} may be attached to each table column, and by that to restrict the set of instances that appear in the data table to those that have certain property values. For example, by applying a filter on the \rdftxt{birthPlace} property, the user may view only those philosophers who were born in \e{Vienna}. Note that by applying data filters, the set $S$ that is captured by the pane is left unchanged. If we want to change our focus of exploration we may ask \elinda to open a new pane that is associated with $S_f$---the set $S$ after applying the filters (\e{filter expansion}). The set $S_f$ may be explored in a similar manner, using all available expansions that will now operate on a narrowed set.

\subsection{Exploring connections with objects}
A third kind of bar chart is available within the \e{Connections} tab and allows to explore connections between $S$ to other sets of instances (now operate as objects) via selected properties. This chart materializes the \e{object expansion}.
The chart provides more insights on $S$ and also the opportunity to \e{switch} an instance set, from $S$ to $O_sp$ (objects connected to $S$ via property $p$). As an example, consider the property \rdftxt{influencedBy} that appears within instances of \rdftxt{Philosopher}. Selecting that property and switching to the \e{Connections} tab opens a bar chart with objects that are connected to $S$ via the property \rdftxt{influencedBy}. The objects are distributed by their type, where each bar represents a set of objects of a given type. One of the bars shown is \rdftxt{Scientist}, denoting the set of scientists who have influenced philosophers. Also here, the user may continue with the exploration on the new set $O_sp$ by clicking a bar. A new pane is opened now focusing on the aforementioned set of scientists. Note that from now on the different expansions will operate on this narrowed set and not on all instances of type \rdftxt{Scientist}.
	\begin{figure}
	\centering
	\input{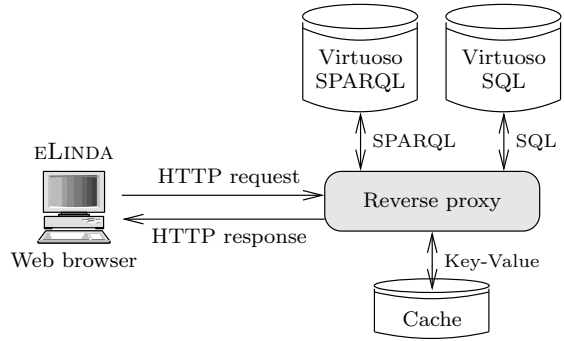}
	\caption{Basic system architecture}
	\label{architecture}
\end{figure}

\section{System Architecture}\label{sec:architecture}
The architecture design of \elinda is driven primarily by the
requirement of \e{responsiveness}, which means that expansions should
happen instantly, preferably in tens to hundreds of milliseconds. Each
exploration step in \elinda is realized by sending one or more SPARQL
queries to the endpoint. Achieving responsiveness is challenging, since some of the queries
that are submitted repeatedly are heavy, with runtime up to several
minutes. Naturally, we can take advantage of the fact that the queries
generated by \elinda are of a restricted form. Next, we describe three
techniques that \elinda implements to boost responsiveness.

\partitle{Incremental evaluation} 
\elinda builds the chart of an expansion by computing it on the first
$N$ triples in the RDF graph. It then continues to compute the query
on the next $N$ triples and aggregates the results in the frontend. It
continues for $k$ steps, or until the full chart is computed.  In the
current implementation, the parameters $N$ and $k$ are determined by
an administrator's configuration. This method provides \elinda with
effective latency for user interaction. Moreover, it works well on
remote servers in the \e{compatibility mode}, as we explain later in
this section.

\partitle{\elinda HVS} \elinda detects heavy queries and 
saves their results in a key-value store called heavy query 
store (HVS) on the \elinda endpoint. For each query to the \elinda 
enpoint, the system first checks if the HVS encountered it before 
and determined it to be heavy. If so, use the result from the HVS, 
otherwise route it to the 
Virtuoso\footnote{\url{https://virtuoso.openlinksw.com/}} endpoint.
\elinda backend measures the run time of the routed queries. 
Queries with runtime bigger than one second are considered heavy and
saved in the HVS. The HVS is cleared on any updated to the \elinda 
knowledge bases.

\partitle{\elinda decomposer} To offer swift accurate results in case
the result is not in the HVS, we provide a specialized \elinda endpoint.  
Our \elinda endpoint contains mirrors of the common knowledge bases, 
such as DBpedia and YAGO, in a Virtuoso SPARQL database. Our system
contains specialized indexes to accelerate heavy queries. \elinda 
detects heavy queries are sent to the \elinda backend and map the SPARQL
queries to a decomposition of SQL queries that utilizes the indexes and 
prevents heavy and redundant SPARQL computations. 
Unlike the \elinda HVS, the \elinda decomposer can be used for \e{all} 
property expansion queries. As an example, on DBPedia the query for the
outgoing property expansion is the following.
\begin{verbatim}
SELECT ?p COUNT(?p) AS ?count SUM(?sp) AS ?sp 
 FROM {SELECT ?s ?p count(*) AS ?sp 
    FROM {?s a owl:Thing. ?s ?p ?o.} 
    GROUP BY ?s ?p} GROUP BY ?p
\end{verbatim}
The first subquery computes the subject, property and count of all the
triples with \rdftxt{owl:Thing} subjects. This query includes a
complex join with hundreds of millions of tuples as an intermediate result,
which delays the response. \elinda decomposer detects that this subquery can
utilize the specialized indexes and decompose the query accordingly.
Clearly, the \elinda decomposer can be used for subclasses of
\rdftxt{owl:Thing}.

\begin{figure}
	\centering
	\includegraphics[width=\columnwidth]{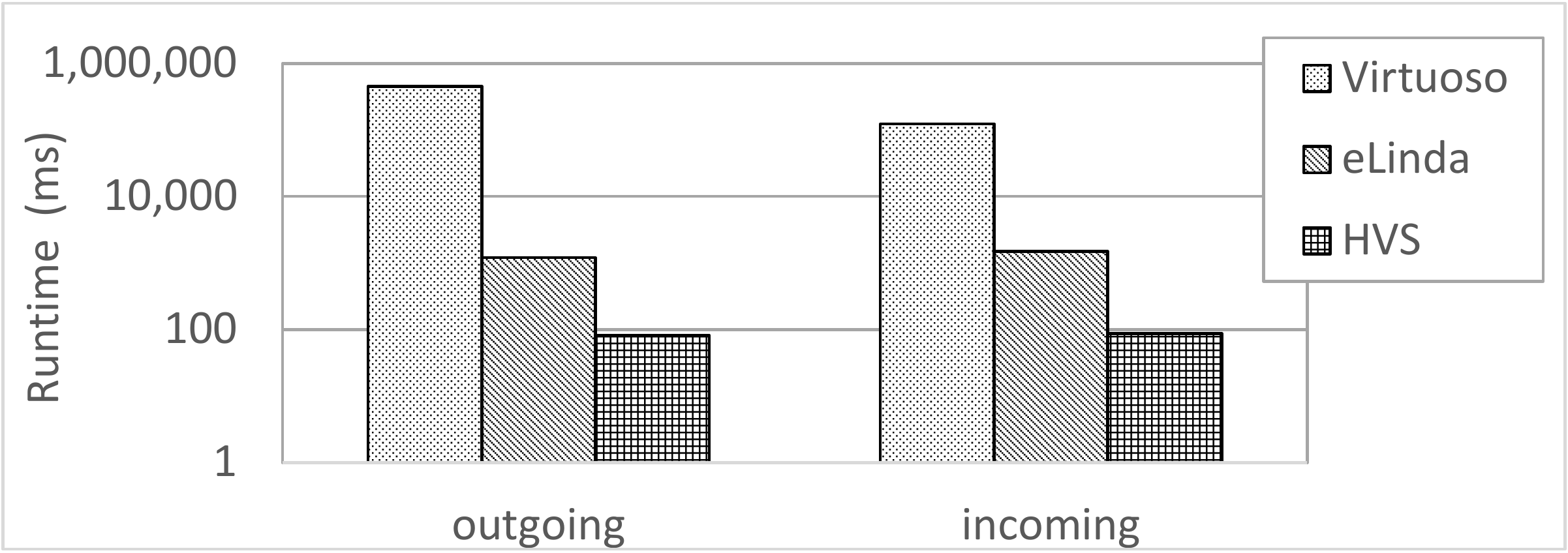}
	\caption{Running times of level-zero property expansions over
		different store configurations}
	\label{queryByStore}
\end{figure}

Figure~\ref{queryByStore} shows the performance of the slowest and
most commonly used queries by \elinda. These queries construct the bar
charts of the outgoing and incoming property expansion in the first
level. The runtime of these queries on the Virtuoso SPARQL endpoint is
454 and 124 seconds for the outgoing and incoming bar charts,
respectively. On the \elinda{} endpoint that uses the \elinda decomposer, 
the queries take 1.5 and 1.2 seconds, respectively. 
In case the results are already in the \elinda HVS, the queries runtime is
around 80 milliseconds. 

\subsubsection*{Remote Compatibility}
Nowadays, data sources often offer online API endpoints that are
constantly evolving~\cite{journals/corr/BikakisS16}. In our
architecture design we also set the goal of allowing the user to apply
\elinda to the exploration of such sources, even if we have no access
to the actual RDF graph and cannot execute any preprocessing.  Hence,
we also allow \elinda to work with a remote Virtuoso
endpoint\footnote{We use AJAX communication with the Virtuoso server
	via its HTTP/JSON SPARQL interface.} that can be configured in the
setting form (described in Section~\ref{sec:ui}) by merely specifying
the endpoint URL. Naturally, in this mode responsiveness is lower than
the above local mode. Yet, the aforementioned incremental evaluation
is applicable (and applied) even in the remote mode, allowing for
effective latency.

	\section{Demonstration Scenarios}
During the demonstration, participants will explore several RDF
datasets such as \dbpedia and
LinkedGeoData\footnote{\url{http://www.linkedgeodata.org/}} with
\elinda. Several kinds of explorations will be exercised. The first
kind will tackle the task of understanding a large and unfamiliar
dataset. Example scenarios are ``examine the bar chart showing the
first-level classes of the dataset,'' and ``analyze the twenty most
significant properties of the largest class in the dataset.''
Additional scenarios will look into sophisticated exploration paths
such as ``the types of people that influenced philosophers.''  A
second kind of exploration will demonstrate the performance issue
elaborated in Section~\ref{sec:architecture}. The participants will be
presented with explorations that entail heavy queries, and with the
discussed solutions turned on and off. Finally, a third kind of
explorations will illustrate how \elinda can be used to detect
\e{erroneous} data such as ``people who are indicated to be born in
resources of type food.''

	{\small
		\bibliographystyle{abbrv}
		\bibliography{ref} 
	}
	
\end{document}